# Human-in-the-loop Extraction of Interpretable Concepts in Deep Learning Models

Zhenge Zhao, Panpan Xu, Carlos Scheidegger, Liu Ren

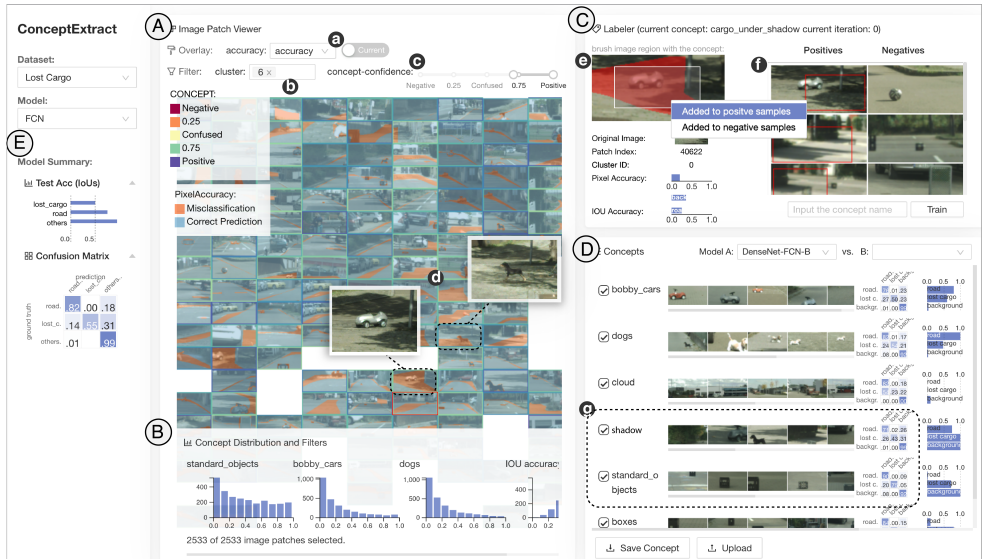

Fig. 1. ConceptExtract supports human-in-the-loop visual concept learning and uses the extracted visual concepts for fine-grained model interpretation, diagnostics and comparison. (A) an image patch view displays small patches or super-pixels (Fig. 4) segmented from the original images with informative overlays to facilitate interactive visual concept extraction. It displays visually similar patches in close proximity without overlap. Here, the overlay shows pixel-wise prediction accuracy for an image segmentation model for road scene understanding, and the border color of each image patch encodes a concept confidence score; (B) a cross-filter panel enable filtering patches based on the concept confidence scores and other statistics; (C) an active learning labeler panel allows easy specification of positive and negative samples for training models; (D) a model analysis panel using the learned visual concepts (*shadow*, *cloud*) for fine-grained model interpretation, diagnostics and comparison. (E) a model summary panel showing a summary of the target model performance. This figure shows ConceptExtract being used to analyze an image segmentation model for road scene understanding in autonomous driving scenarios. Please refer to Section 7.2 for more detail.

**Abstract**—The interpretation of deep neural networks (DNNs) has become a key topic as more and more people apply them to solve various problems and making critical decisions. Concept-based explanations have recently become a popular approach for post-hoc interpretation of DNNs. However, identifying human-understandable visual concepts that affect model decisions is a challenging task that is not easily addressed with automatic approaches. We present a novel human-in-the-loop approach to generate user-defined concepts for model interpretation and diagnostics. Central to our proposal is the use of active learning, where human knowledge and feedback are combined to train a concept extractor with very little human labeling effort. We integrate this process into an interactive system, ConceptExtract. Through two case studies, we show how our approach helps analyze model behavior and extract human-friendly concepts for different machine learning tasks and datasets and how to use these concepts to understand the predictions, compare model performance and make suggestions for model refinement. Quantitative experiments show that our active learning approach can accurately extract meaningful visual concepts. More importantly, by identifying visual concepts that negatively affect model performance, we develop the corresponding data augmentation strategy that consistently improves model performance.

---

- *Zhenge Zhao is with the University of Arizona. E-mail: zhengezhao@email.arizona.edu.*
- *Panpan Xu is with Amazon AWS AI. Email: xupanpan@amazon.com.*
- *Carlos Scheidegger is with the University of Arizona. E-mail: cscheid@cs.arizona.edu.*
- *Ren Liu is with Bosch Research North America. E-mail: Liu.Ren@us.bosch.com.*

## 1 INTRODUCTION

Deep neural networks have achieved state-of-the-art performance in many challenging computer vision tasks and are being widely adopted in many real-world application scenarios such as autonomous driving. As a result, recent emphasis on deep learning models has moved from model accuracy alone towards issues such as model interpretability. The machine learning community has realized the necessity of making the models more understandable, especially since these models can easily have hundreds of millions of parameters with highly non-linear transformations. First of all, the model/application developers might want to scrutinize the decisions made by machine learning models and use them more responsibly [21]. If the model developers can understand the weaknesses of their AI models, they could minimize the potential errors or biases of training data in real-world applications.

This is orthogonal to model accuracy: a recent study by Bansal et al. [4] shows that increasing AI accuracy may not bring the same improvements for performance if the human cannot develop insights into the AI system. Secondly, **improving model interpretability can facilitate model refinement**.

For instance, when designing AIs for autonomous driving, the detection of unexpected road hazards such as lost cargo [33] is a typical image segmentation task in computer vision. When model developers train a neural network like Fully Convolutional Network (FCN) [27] or DeepLabV3 [9] for lost-cargo detection, the accuracy is relatively low and the developers have difficulties in finding potential root causes [25], which could be the lighting conditions on the road, the visual features of the lost-cargo objects themselves or others. Identifying such potential root causes can help develop mitigation strategies (e.g. applying appropriate data augmentations) to further improve the model, and model interpretation is the key to discover such root causes.

To tackle the issue of interpretability in neural networks, many techniques [1, 35] have been proposed to help people understand model predictions. TCAV (Testing with Concept Activation Vectors) and the follow-up work ACE aim to understand what signals the model uses for predicting different image labels [13, 23]. They generate a measure of importance of a *visual concept* (e.g. wheel, glass) for a prediction (e.g. predicted as a car) in a trained model. However, the concepts generated by automatic clustering methods may not match human concepts.

**In other words, such methods cannot guarantee that image patches which are relatively close and gathered in a latent space are semantically meaningful to humans as a concept. This mismatch provides the inspiration for our work.** We propose a visual analytics framework to integrate human knowledge in the visual concept extraction process and use the identified concepts to analyze potential causes of model errors and develop mitigation strategies. **Specifically, we propose a novel combination of an active learning process with a user interface expressly designed for fast labeling of images to train a concept extractor network** that identifies patches containing a common concept. Our system ConceptExtract enables users to explore image patches, control the active learning process and use the resulting concepts for model comparison and diagnosis. We present example usage scenarios for different datasets and machine learning tasks, including image classification for ImageNet [11] and image segmentation for the lost cargo challenge [33]. We analyze a variety of neural network architectures, including ResNet [15], VGG [40], FCN [27] and DeepLabV3 [9], demonstrating the generality of our proposed approach. Using ConceptExtract, users can extract semantically meaningful concepts, provide concept-based explanations for different machine learning models and compare them. Our quantitative evaluation (presented in Section 8) shows this approach produces concept extractors accurately and more efficiently than random labeling or traditional active learning approaches. Furthermore, we show the validity of the concepts we extract by following up the concept extraction procedure with an associated data augmentation strategy that improves the performance of model under analysis.

In summary, we contribute:

- A novel visual analytics framework supporting a human-in-the-loop, active learning based approach to extract visual concepts for model interpretation, as well as identifying visual concepts that negatively affect model performance (Section 4);

- A prototype system implementing our proposed human-in-the-loop workflow, featuring scalable image patch exploration, visual cues and interactive filters for active learning and a rich set of model diagnostics and comparative analysis visualizations (Section 5);

- Two case studies and quantitative experiments demonstrating the value of using ConceptExtract for diverse machine learning tasks and datasets (Section 7).

- Quantitative experiments that show ConceptExtract produces concepts faster than traditional active learning, and that these concepts can help develop data augmentation strategies for model performance improvement (Section 8).

## 2 BACKGROUND

### 2.1 Deep Neural Networks

In this paper, we use neural networks whose first layer has as many units as there are pixels in the input image. To exploit spatial locality, current deep neural networks (DNNs) use *convolutional layers*, typically followed by nonlinear activation functions. After a sequence of such layers, a fully-connected layer is usually present before the model output. This basic setup can be used in various tasks by assembling different layers; each potential configuration is called an *architecture*.

### 2.2 Deep Embeddings

To obtain image patches that potentially contain the same concept, we need some approach to measure image similarity. However, direct pixel difference measurements fail to take into account misalignment, distortions, lighting changes, and so on. To solve this problem, we use *deep embeddings* as a representation of the image patches. As an image is passed as an input through a DNN model, the output after each hidden layer is an embedding in that latent space. These deep embeddings provide hints for the model to distinguish different images. Previous work shows that euclidean distance[1] in the latent space is an effective perceptual similarity metric [54]. In this paper, we resize inputs to match the architecture's first layer and choose the embeddings from a low-dimensional layer as the latent representation. (In this paper, we use deep embedding interchangeably with latent representation.)

### 2.3 Active Learning

Active learning is a semi-supervised machine learning method where the learning algorithm can interactively query a user for labeling instances. Instead of manually labeling all the unlabeled instances, active learning makes a priority to label the data that have the highest impact on training the model. This method is widely used in training neural networks [36, 48, 50]. Commonly used prioritizing methods include model confidence, margin sampling, and entropy [38]. Once an approach has been chosen to prioritize the labeling, this process can be iteratively repeated: a small subset of data with the highest prioritization scores will be presented to the user to assign labels. After that, the DNN can be trained on the manually labeled data. Once the model has been trained, the unlabeled data points can be run through the model to update their prioritization scores, which significantly reduces the overall labeling burden. In this paper, we show that allowing a user to pick from a set of carefully laid-out images produces a more efficient sequence of training models than is possible with pure sequential active learning.

### 2.4 Concept Annotations

In image classification, all possible categories are assumed to be known to the model, and images are typically assumed to belong to a single class. However, an image may be complex (for example, it can contain various objects and visual patterns). We refer to these "potential" labels of these objects as *Concept Annotations*. They are different from the classification labels, and an image may admit multiple concept annotations. Concept annotations are not used in training the network for the task, but they can provide the grounding necessary for model explanations.

## 3 RELATED WORK

The goal of providing a human level of understanding of a deep learning model now drives an entire subfield of machine learning research. While some work focuses on building inherently explainable models [2, 8, 10, 17, 29, 52, 53] to achieve interpretability, we in this project focus on **post-hoc explanations**, interpreting models that were trained without any consideration to interpretability.

---

[1] The other common metric option could be cosine similarity.

**Post-hoc Explanations** Saliency methods form a popular class of tools that provide localized explanations for each data sample by calculating the importance of each input feature (typically pixels). Backpropagation methods like Gradients [39], DeconvNets [51] and Guided Backpropagation [43] compute the gradient of the network's prediction with respect to the input. LRP [3] redistributes the relevance of each neuron through additive functions that conserve a measure of total importance from layer to layer. VisualBackProp [5] uses deconvolutions to arrive at a pixel-based importance metric. While saliency methods can illustrate the attention of deep learning models, the resulting heatmap itself can be hard to interpret and assess. The study by Adebayo et al. [1] shows that some of these popular methods fail the sanity check. Data perturbation methods [12, 32, 51] use small prediction changes to generate interpretations. In particular, LIME [34] and SHAP [28] change the input in a controlled fashion and observe the effect on the output. Slack et al. demonstrate that LIME and SHAP can be easily fooled by crafting adversarial classifiers, showing a potential general weakness of perturbation methods [41]. While most saliency methods focus on a single data sample, our system adopted concept-based explanations to provide a higher level interpretation that aligns better with human knowledge.

**Inherently interpretable models** One way to achieve interpretability is by building models which are inherently explainable. Mimic learning [2] replaces the deep neural networks with models that are easier to explain. Choi et al. [10] propose RETAIN, a sequence model with an attention mechanism for highlighting the most meaningful visits of the patients for their diagnosis results. Zhang et al. [52, 53] propose methods that enforce the interpretability of high-level image filters through disentangled representations. ProtoPNet [8] is a deep network architecture that picks out essential patterns in the image and generates a prediction by comparing those patterns to typical classes it has seen before. Ming et al. [29] combines prototype learning with deep sequence models to achieve interpretability. TED [17] suggests that while training a model, the objectives should combine both the explanations and the labels. However, currently how to combine them with popular architectures remains unsolved, and training such models are often time-consuming. Our approach focuses on diagnosing trained model and doesn't need to retrain the original model during the analysis.

**Visualizing latent spaces** A latent space in deep learning is a reduced-dimensionality vector space of a hidden layer. The neural network compresses the input and forms a new low-dimensional representation with interesting properties [45, 49]. Here, we employ latent space techniques to build concept extraction models and provide a good spatial arrangement for users to select images to label. Recently, some interactive visual systems have emerged to facilitate the exploration of the latent space. Spinner et al. [42] propose an interactive visualization for comparing two different models by exploring their latent spaces. Liu et al. [26] propose LSC (Latent Space Cartography), a comprehensive system for mapping and comparing meaningful semantic dimensions within latent space. SMILY [7] is an interactive system to help pathologists search for similar medical images of patients based on the users' preferences. We use latent spaces indirectly as the input for our concept extraction networks and use latent spaces to drive the layout of the image patch view. More than that, by utilizing different visualization techniques, ConceptExtract can help the users easily explore the dataset and summarize their findings.

## 4 TASKS AND WORKFLOW

### 4.1 Task Analysis

Model developers encounter different problems while diagnosing their model to make improvements. Developers want to understand predictions and find the leading causes of a specific result. For example, if a classification model predicts an image as a "fish", is it because it recognizes the fish body, or is it using contextual cues in the image such as a human holding it or a container carrying the fish? This same question has been studied in Summit [20] as well. Another main concern is the identification of systematic causes of misclassification. When developing a semantic segmentation model for analyzing road scenes in an autonomous driving application, developers find that a dog is incorrectly detected under a tree shadow. Is this just a coincidence or a common phenomenon happening across the entire dataset? Answering these questions not only help the developers better understand and anticipate the model behavior, but also helps them develop effective strategies to refine the model and improve its performance.

With ConceptExtract, we seek to give model developers a visual analytic system so they can interpret, diagnose and compare deep learning models with human-friendly, concept-based explanations. Concretely, based on previous discussions in visual analytics for deep learning [19] and the requisite expertise of the co-authors from past experience, we start by identifying a set of analytic tasks to be supported in the system.

**T1: Summarize model behavior.** The system should provide a summary of the model to the developers to start with. Deep learning models can have different performance metrics depending on the task, e.g. precision in image classification model and IoU (Intersection over Union) accuracy in semantic segmentation models; prompt access to these measures is a requirement.

**T2: Browse and explore image patches/super-pixels.** It is challenging for users even to know what visual concepts exist in the data. Since each dataset potentially contains many concepts, it is important for the user to be able to extract visual concepts that highly influence model decision. The system, therefore, needs to provide an overview of the image patches with a good layout strategy, as well as also provide a set of filters to help users quickly identify interesting data samples and decide which image patches to study first.

**T3: Train and evaluate concept extraction models.** Since no ground truth labels exist for visual concepts, and it is infeasible for users to manually label a large number of images, we propose using a separate concept extraction active learning loop to efficiently derive a set of image patches containing a visual concept. The system should involve users' human knowledge and give them the flexibility to choose and customize any potential concept they recognize in the image patches. It should also provide methods for the user to evaluate whether the model has sufficiently learnt the visual concept.

**T4: Analyze how visual concepts affect model decisions.** After extracting human-friendly visual concepts, the system should support using them to understand model behavior. The system should help users systematically analyze how important the visual concepts are for predicting different classes and analyze how the presence of different visual concepts in images affects model performance (e.g. shadow prevents detection of objects on the road).

**T5: Compare different models.** In addition to investigating the target model, the system should further support using the visual concepts extracted for fine-grained model comparison, esp. how the performance of the models differ on images containing different visual concepts. This helps reveal the strength and weaknesses of different models.

### 4.2 Workflow

To integrate the tasks described above, we present a workflow to help guide the user through the analysis steps in ConceptExtract. We will refer to Fig. 2 throughout the section. For the specific settings of the workflow in real-world applications, please refer to Section 7.1 and Section 7.2 for more details.

The workflow starts with a **preprocessing** stage (left portion) with the available image data and the target model. In the preprocessing stage, the original images are segmented into patches using fixed window sizes or super-pixel segmentation algorithms [47]. The image patches/super-pixels are then resized to the input scale and fed to the target model. Their latent representations are extracted at a selected layer in the target model for visual concept learning. The **visual concept learning** stage (center portion) uses concept extractor networks on top of the latent representations to learn human-understandable concepts and retrieve image patches containing those concepts for model analysis (**T3**). Individual networks with the same architecture but different weights are trained to recognize different visual concepts through an active learning process.

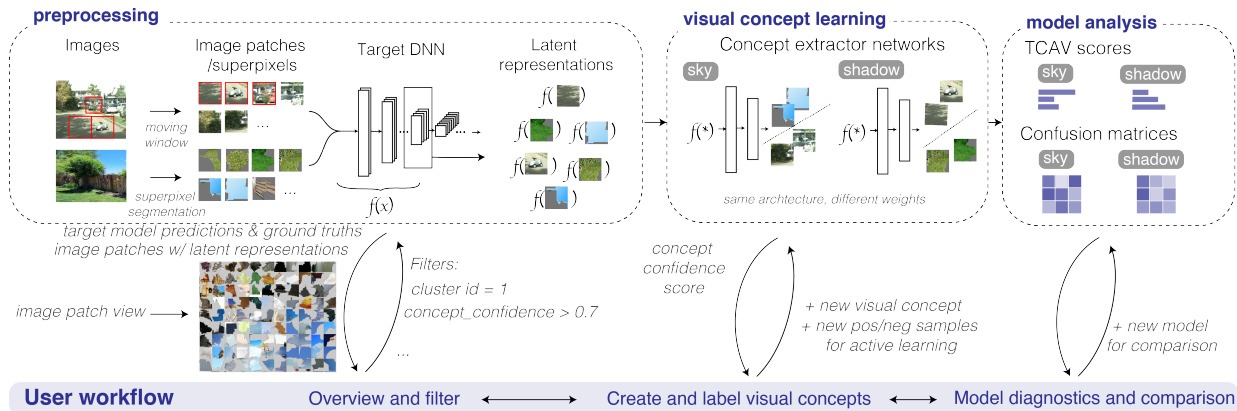

Fig. 2. The workflow of ConceptExtract. In the *preprocessing* stage, the images are segmented into image patches or super-pixels. The latent representations of these image patches/super-pixels are extracted from a selected layer in the target model. The visualization interface layout the image patches such that similar patches are spatially close. Users can easily identify and create new visual concepts and overlay data such as target model misclassifications to focus on problematic cases. In the *visual concept learning* stage, we utilize concept extractor networks to retrieve image patches containing the same concept (*sky* and *shadow* in the figure). The concept extractor networks take the latent presentations of the image patches as inputs and output *concept confidence scores* in [0, 1]. We employ a visualization assisted active learning process to train the concept extractor networks. The learned visual concepts are used in the *model analysis* stage for model interpretation and comparison with visualizations such as TCAV scores charts and confusion matrices.

To help users create meaningful novel visual concepts, the system provides an overview of the image patches and projects them in a way such that visually similar image patches are close to each other. The user can also interactively overlay a variety of information on top of the image patches such as accuracy, ground-truth and predicted labels to prioritize looking for visual concepts that affect model performance (**T2**).

To support effective novel visual concept learning and reduce user labeling effort in the active learning process, we propose a hybrid approach that tightly couples visualization and computational techniques (**T3**). For each image patch, the concept extractor network produces a *concept confidence score*. The concept confidence score ranges from 0 to 1, where 0 is for confidently negative (the image patch does not contain the visual concept), 1 is for confidently positive (the image patch must contain the visual concept), and 0.5 is for not sure. The system visualizes the concept confidence score and supports interactive data filtering based on it to help the users prioritize labeling more informative examples for model training. **In particular, we found that labeling hard negative samples [37], which are image patches confidently but wrongly classified, can greatly facilitate the training process**. The user can also filter the image patches with the most confident predictions to verify if the concept extractor has been sufficiently trained to recognize visual concepts that align with human-knowledge. To further reduce user effort and recognize novel visual concepts with very few labeled examples provided by the user, we also use a data augmentation strategy [18] which has been proven to be effective in similar scenarios such as few-shot learning or zero-shot learning [24]. The data augmentation method selects each labeled image patch, randomly applies two categories of augmentation policies: (1) shape policies like shearing, flipping, rotating, and (2) color policies like gray-scaling and blurring.

After obtaining a set of visual concepts and the corresponding image patches, the user can move to the **model analysis** stage (right portion) and perform model interpretation, diagnostics and comparison (**T4** and **T5**) using TCAV scores and confusion matrices. The visualization shows fine-grained analysis, including how each visual concept affects the model's and how the model performances differ on images containing different visual concepts.

## 5 SYSTEM DESCRIPTION

As shown in Fig. 1, the visual interface consists of a set of visualization modules to display information like model summary (**T1**), visual concept images, and a series of interactions to support image patch explorations and the active learning process for training the concept extractor network. We will discuss the main components in ConceptExtract and how the these components together support the workflow ( Fig. 2 ).

### 5.1 The Image Patch View

The image patch view ( Fig. 1 (A)) provides an overview of the image patches to help the user quickly explore the data collections and identify interesting visual concepts (**T2**). We apply t-SNE [46] to the image patches' latent representations to get a 2D layout. Since directly plotting the image patches according to the projected coordinates will result in severe visual clutter, we use a de-cluttering algorithm to layout the image patches in non-overlapping grids while still keep visually similar image patches close to each other. Specifically, we partition the canvas area into grids with identical size rectangles. Then we randomize the image patch sequence. For each image patch, we find the grid cell containing the 2D coordinates. If the grid is empty, we plot the image patch on the grid. If the grid is already occupied, the layout algorithm will search for the nearest neighbor grids to fill. When no empty grid is available on the screen, the image patch will be hidden temporarily. Navigation operations like zooming in will increase the number of grid cells available. When a different scale is reached, we re-plot the image patch view to allow more image patches to be displayed on the screen. We bring similar image patches as close as possible through this layout while reducing visual clutter due to overdraw.

A control panel on top of the image patch view allows users to overlay additional information on the image patches as well as filter the data. When the users first explore the data, it is challenging for them even to know where to start their study. The "cluster" filter ( Fig. 1 (b)) gives the user the option to plot only image patches in the selected clusters precomputed using algorithms such as *k*-means. Users can also choose color overlays or border highlights on the image patches to show information such as ground-truth, model predictions and model accuracy ( Fig. 1 (a)). For example in Fig. 1(A), for an image segmentation model, the visualization displays pixel-wise image segmentation accuracy where red indicates the wrong prediction and blue indicates the right prediction. In another example shown in Fig. 4(a), the visualization uses border color to indicate whether the source image of a super-pixel is correctly classified in an image classification model. With different overlays, users can focus on particular image patches to extract the relevant visual concept. For example, the user is usually interested in image patches related to wrong predictions ( Fig. 1 (d)) and extracting visual concepts from those image patches could better benefit model diagnostics.

As a crucial component in the active learning process, the control panel also has a range slider ( Fig. 1 (c)) to help users efficiently filter the data based on the *concept confidence score* of each image patch for the concept currently being trained. The user can also draw the concept confidence score as the border color of the image patches in a diverging color scheme, as shown in Fig. 1.

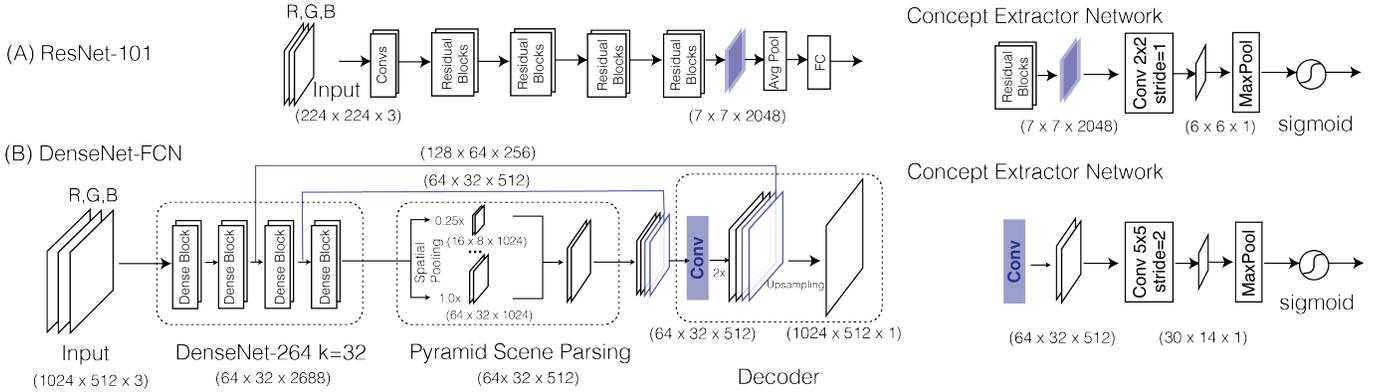

Fig. 3. We apply ConceptExtract to analyze image classification and segmentation DNNs: (A) ResNet-101 [15] trained on ImageNet (B) a Fully Convolutional Network (FCN) [27] with a DenseNet encoder [22] and a Pyramid Scene Parsing (PSP) [55] structure trained for road image segmentation. We extract deep embeddings from the smallest hidden layers available. The chosen layers are highlighted in the architecture diagrams. The concept extractor network used for each architecture is shown on the right.

### 5.2 The Training View

The training view ( Fig. 1 (C)) provides a frontend to control the active learning process (**T3**). It contains two parts: a patch details & interaction area (( Fig. 1 (e)) for the user to assign concept labels and a training samples list ( Fig. 1 (f)) for showing selected images and their training status. The selected image patch from the image patch view will be magnified, and the related information will be presented such as the source of the image patch. The user can directly add any patch that doesn't contain the concept into the negative training set by selecting on the context menu. To add positive samples, the user either crops a rectangle on the image that contains the concept and discards the rest of the pixels, or directly selects the whole image patch/super-pixel as a positive sample. All the selected positive and negative samples will be displayed in the training samples list ( Fig. 1 (f)). Concepts can be named and saved for use in future sessions. While the active learning network is trained, the user can continue adding different image patches into the training set, or end that training stage, save the concept extractor network and the retrieved images containing the concept.

### 5.3 The Model Analysis View Using Visual Concepts

The Model Analysis View uses the learned visual concepts to support fine-grained model interpretation, diagnostics and comparison ( Fig. 1(D) ). After a user completes a new concept extractor's training process, ConceptExtract shows the record of this concept in this area, including the concept name and the image patches with the highest confidence scores. A barchart shows TCAV scores for each visual concept, and the length of each bar indicates the importance of this concept for predicting a specific class (**T4**). To gauge a potential weakness of the model being analyzed with respect to the concepts, we choose for each concept the top 50 image patches based on the concept confidence score, find the original images of these image patches and compare predictions of our target model with the ground-truth using a confusion matrix (**T4**). Each row in the confusion matrix represents the ground truth class, and each column represents the predicted class. The values on the matrix diagonal show the proportion of the data samples correctly classified in each class. We use a sequential colormap to encode the proportion ranging from 0 to 1. With the confusion matrices, the user can analyze whether the presence of a certain visual concept in the image leads to more model errors. An example is shown in Fig. 1(g), where the model being analyzed has worse performance on images containing the *shadow* concept.

We can also use the learned visual concepts to to compare different models visually. For the two selected models from a list, we compute their confusion matrices for each of the visual concepts and then directly calculate the difference between them. The differences are displayed using a diverging colormap, where red indicates negative values and blue indicates positive values in the matrix. If a second model has better performance than the first one, the diagonal entries should show more positive values (blues) in the matrix and vice versa. For example, in Fig. 6(b) we compare DenseNet to ResNet on images containing the visual concept *sky*. Since there are more red colored entries on the diagonal, we can conclude that DenseNet has worse performance on this set of images. Such comparison reveals the strength and weaknesses of each model and helps identify opportunities to use model ensembles to improve prediction accuracy.

### 5.4 Other Views

The model summary view ( Fig. 1 (E)) shows basic information like the datasets and the model types. We use both bar charts and confusion matrices to show model performance on different classes (**T1**). A cross-filter view ( Fig. 1 (B)) shows the distribution of image patches based on different features, supporting quick retrieval and comparisons (**T4**). In this view, each image patch could be treated as a multivariate data sample, including variables like prediction accuracy and concept confidence scores for the existing concept extractors. A barchart is displayed for each of these variables. To help the user quickly identify an interesting target and generate new facts, the crossfilter view is also connected with the image patch view. Only the selected image patches in the crossfilter will be plotted in the image patch view. These concept filters can help the user quickly identify confident or confused image patches for different concepts. It is particularly useful when the user has trained multiple visual concepts and would like to study how the learned concepts correlate with each other.

## 6 SYSTEM IMPLEMENTATION

Our system design separates the frontend for data visualization and the backend for data storage and active learning. For the backend of the system, we use Pytorch [31] to implement the target machine learning models including DenseNet-FCN (We use the implementation in this repository: https://github.com/sagieppel/Fully-convolutional-neural-network-FCN-for-semantic-segmentation-with-pytorch.) and ResNet-101 [16], as well as other models for comparison including DeepLabV3+ [9], DenseNet [22], VGG [40]. We also use Pytorch to implement and train the concept extractor networks. To extract visual concepts, all the images are segmented into small image patches or super-pixels of different sizes. We use scikit-image (https://scikit-image.org/) for super-pixel extraction. The image patches or super-pixels are then scaled to the same size as the input of the target model. By running them through the target model, we extract and save the latent representation of these image patches (or super-pixels) at the selected layer. All image patches, along with their latent representations, ground-truth labels, predicted labels and (per-pixel) accuracy are stored in the backend system as binary files in the file system. The application web server is implemented with Flask [14]. For the frontend design, we mainly rely on two JavaScript libraries, React and D3 [6] and draw on both SVG and HTML5 Canvas for better performance.

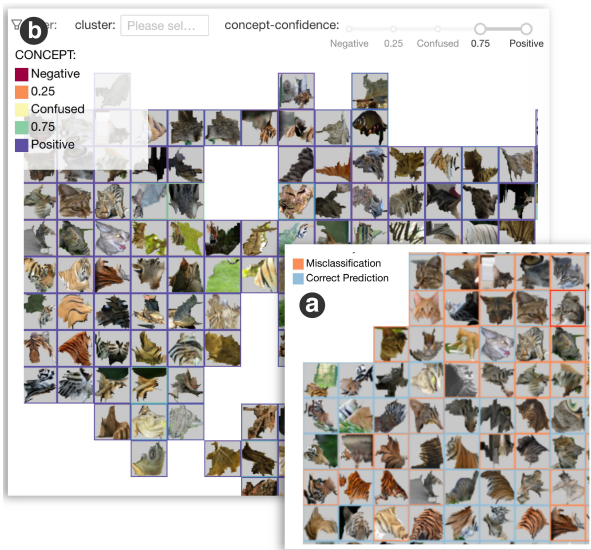

Fig. 4. (a) The user identifies that many stripe patterns are associated with erroneous predictions (b) Through active learning the concept extractor network is able to accurately retrieve large amount of super-pixels containing the strip patterns. The stripe concept will be used for further model analysis.

## 7 EXAMPLE USAGE SCENARIOS

We demonstrate ConceptExtract in example usage scenarios on two different perception models for image classification and semantic segmentation tasks. By utilizing our system, interesting visual concepts are revealed in the models and the datasets. We further demonstrate how to use these concepts to help model interpretation, model comparison, and developing mitigation strategies for model performance improvement.

### 7.1 ResNet-101 for Image Classification

In this section, we demonstrate the application of ConceptExtract in analyzing an image classification model trained on ImageNet data. We show that the system can help extract semantically meaningful visual concepts. ImageNet Large-Scale Visual Recognition Challenge (ILSVRC) is a famous competition which has been held annually since 2010. ILSVRC uses a subset of ImageNet [11] with roughly 1.4 million images belonging to 1000 categories. For simplicity, we choose a subset of ImageNet with 10 classes which are the 10 worst performing classes of the model (*tench*, *tabby*, *tiger cat*, *tiger*, *barracouta*, *balloon*, *castle*, *church*, *parachute*, *vault*). Also, most of the wrong predictions of these 10 classes made by the model are among the same 10 classes, for example, tench is often misclassified as another type of fish, tabby. In each class, we randomly sampled 200 images for analysis, which we think can be processed quickly by our system while generating enough concept patches. We analyze the pre-trained ResNet-101 [16] model from PyTorch with top-1 error 22.63% and top-5 error 6.44%. To extract concepts of different resolutions from these classes, we use the Quickshift algorithm [47] to compute a hierarchical segmentation of the images on multiple scales. For each image, Quickshift generates about 12 super-pixels. After that, we resize these super-pixels to the original input size of the model and pad the empty pixels with neutral gray. We choose the final convolutional layer ( Fig. 3(A) ) with dimensions 2048 × 7 × 7 to extract the latent representation of each super-pixel.

**Concept Extractor Network Setup** Our concept extractor network ( Fig. 3(A) ) contains only two layers on top of the latent representation extracted from ResNet-101: one convolutional layer and one max pooling layer. A sigmoid function is applied after the max pooling layer to obtain a ***concept confidence score*** between 0 and 1 to predict whether the super-pixel contains the specified visual concept or not. This simple architecture is accurate enough for identifying concept images (see Section 8 for details), and training such a network will not take the user a lot of waiting time. For each stage, the neural net is trained until the validation loss does not decrease. We observe that in most cases the training stops at around 10 epochs. Note that only the weights in the concept extractor network are updated (and not the task model under analysis). The ***concept confidence score***s are sent to the frontend after the training is done. They are displayed in the image patch view and can also be used to filter the image patches to find the most informative examples to label. To obtain more training data, we generate 200 images for both positive and negative training sets by data augmentation. The 200 images are generated by first randomly sampling from the available training data and then applying the data augmentation strategy.

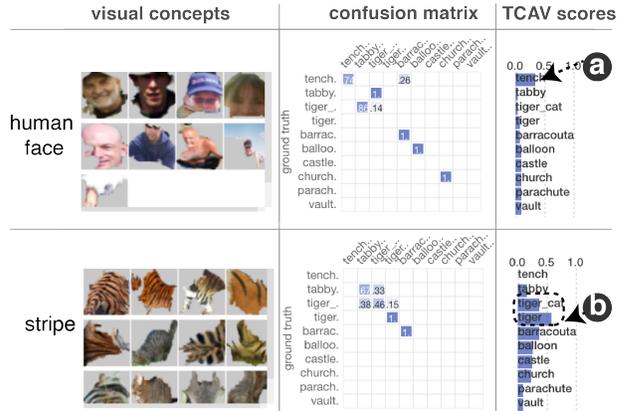

Fig. 5. (a) TCAV score shows that the human face is surprisingly important for identifying the class *tench*, a type of fish. (b) stripes is an important visual concept for identifying tiger cat and tiger and is also the main reason that these two classes are often confused with each other.

**Extracting Visual Concepts** The system initially displays all the super-pixels in a compact layout, generated from a t-SNE projection of the latent representation. The layout places semantically-similar super-pixels in clusters that can have associated concepts such as grass and sky. To prioritize finding concepts that affect model performance, the user can overlay predictive accuracy for each super-pixel. For example, one can observe a cluster of super-pixels showing orange and black strip patterns are often misclassified ( Fig. 4(a) ). The user therefore starts creating a new visual concept "orange and black stripe" by adding new labeled examples. After specified 4 to 5 positive and the same number of negative samples, the user can click on the "Train" button in the Labeler to start the first training stage for this concept extractor. The training time depends on the dimension of the latent representation and the GPU configuration. On a machine with a GTX 1070Ti GPU, it typically takes about 50 seconds to train one stage. Based on the returned concept confidence score, the user can use the filter to select more informative examples to label, esp. hard negative samples which are confidently but wrongly classified by the concept extractor. After several iterations, the user finds that almost all the super-pixels filtered with a range of high concept confidence score (e.g. 0.75 1.0) contains orange and black stripe patterns ( Fig. 4(b) ) and all the super-pixels filtered with low and medium concept confidence scores (e.g. 0.0 0.5) do not contain the stripe patterns. Therefore the user can consider the concept extractor network has successfully learned the orange and black strip concept and use it for model analysis and comparison.

The user can continue exploring the image patch viewer and create new visual concepts following a similar process. For each new visual concept, an individual concept extractor network is created and trained. They all have the same architecture as described in Fig. 3(A). For example, she can train four separate concept extractor networks to identify visual concepts, including *human face*, *fish*, *stripes*, and *sky*.

**Model Analysis with TCAV Scores and Confusion Matrix** From the TCAV scores, we can identify that the *human face* concept is highly relevant for predicting the class *tench* ( Fig. 5(a) ), a type of fish. This result aligns with the findings in Summit [20], a previous paper that also analyzes deep neural networks for ImageNet classification. They also found that predicting the class *tench* relies heavily on person related features. Notice that we use ResNet-101, not InceptionV1 [44] as in Summit. This leads to the hypothesis that the data distribution, in-

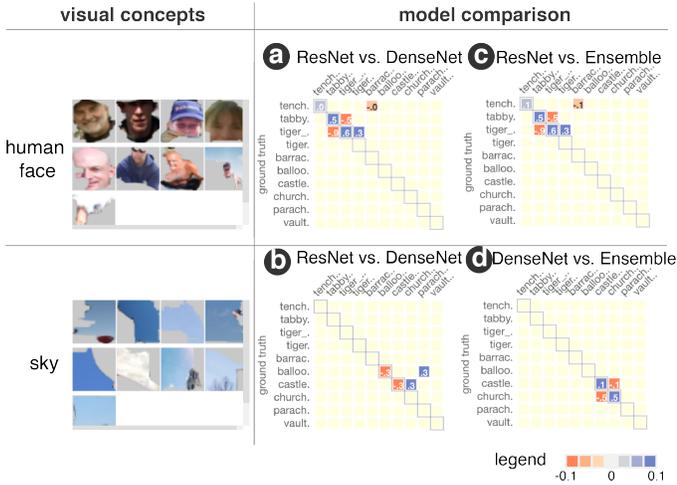

Fig. 6. The confusion matrices show pairwise model comparison, with fine-grained information about which model performs better on images containing a given concept. DenseNet performs better than ResNet on images containing human faces (a) but worse on images containing sky (b). The two models show complementary strength, suggesting that a model ensemble averaging their predictions outperforms both (c and d).

DenseNet-FCN performs semantic image segmentation by predicting a label for each pixel in the image. In this case, each pixel could belong to three different classes, including lost-cargo (obstacles), road, and background. As shown in Fig. 3(B), to extract the latent representations for concept learning, we chose the layer at the beginning of the decoder (dimension: 512 ×32 ×64) for two reasons: (i) the layer encodes both local and global information, (ii) the layer has the most compact size, which will benefit future computation and storage.

For this task, since the model designers want to keep the context of potential concepts, we use rectangle boxes with three different sizes to obtain image patches for extracting concepts instead of segmenting the image into super-pixels. Since there are a large number of image patches (over 4 million), we sampled a subset of them for analysis. Furthermore, since the main task is to detect the lost cargo on the road, we chose all the image patches containing lost cargo (roughly 1000) and sampled around 1000 image patches containing the other two labels: road and background. In all, we have 2533 image patches for concept extraction and visualization.

The lost cargo has two types of pixel annotations: the coarse ones including lost cargo (obstacle), road, and background; the fine ones for distinguishing specific lost cargo objects/obstacles in the images like boxes, balls, and so on. The coarse annotations are used by DenseNet-FCN for training and prediction. To quantitatively evaluate our concept extraction model, we use the fine annotations as groundtruth visual concepts. We pick a concept — dogs and trained the concept classifier for 4 iterations. Ten positive and ten negative images are selected for the initial stage, and for each of the rest stages, four positive and four negative images are added. The results are presented in Fig. 7. The figure plots the precision of the concept extractor when retrieving top-$k$ image patches according to the concept confidence score. For each active learning stage, we can see a significant improvement in the precision of the predictions after the active learning process, especially for the top 50 image patches based on the concept score.

As shown in Fig. 1(A), to prioritize the visual concepts that affect model performance, the user overlays the pixel accuracy of the model prediction on each image patch. While exploring these image patches, the user identifies that sometimes the lost cargo cannot be correctly detected when it is under a tree shadow ( Fig. 1(d)). Is this just a coincidence, or is it happening across the entire dataset? To answer this question, the user creates the visual concept named "shadow". She also starts specifying positive and negative samples for the concept extractor to learn to retrieve similar image patches also containing shadow. The training process also utilizes the data augmentation strategy described in Section 4.2. The data augmentation process generates 200 images for both positive and negative training sets.

The model analysis result of "shadow" is displayed in Fig. 1(g), together with some other concepts. From the confusion matrix, the user can verify that indeed the DenseNet-FCN model performs worse on the images containing "shadow" images compared to images containing other concepts such as standard objects. Meanwhile, the TCAV score indicates that the "shadow" pattern influences the prediction of all three segmentation labels (in the image segmentation model, we consider each pixel as an individual data sample to compute the TCAV score [23]). To validate this hypothesis, we augmented the training set with artificially-generated shadows ( Fig. 8 ). We randomly draw a boundary line across the lost cargo's bounding box. On a random side of the line, we apply a brightness reduction. To make the shadow more realistic, we gradually change the darkness around the boundary with Gaussian blur. As shown in Table 1, the fine-tuned model after augmentation is more accurate. To further verify this strategy's scalability, we also apply the shadow augmentation to another state-of-art model, DeepLabV3+ [9] and we see an improvement for IoU accuracy as well.

For this particular usage scenario, we interviewed and gathered feedback from an industrial expert with 15+ years of experience in computer vision research and has been heavily involved in the development of autonomous driving software. We introduced the main idea of using visual concept learning for model diagnostics and presented a walk-through of the system through a remote video call. He immediately identified that the system has great potential for collecting similar stead of the model architecture, is causing the problem. Since the training data contains a lot of images of person holding *tench*, both models automatically make use of such visual concept to perform classification.

Based on the TCAV scores we can also observe that the three frequency confused classes *tiger cat*, *tiger* and *tabby cat* all uses stripes as a visual concept to perform classification ( Fig. 5(b) ). The confusion matrix shows that on images containing stripes, the model often make mistakes among the three types of feline animals.

**Compare Different Models** The visual concepts extracted can be reused to obtain a fine-grained comparison between different models, which goes beyond simple benchmarks such as overall model accuracy. In particular, we can analyze which model is better at classifying images containing a certain concept. In this example, the user loads another state-of-the-art model DenseNet [22] to compare it with ResNet-101. Based on the confusion matrix the user observes that while DenseNet performs better than ResNet-101 on images containing visual concepts like *human-face*, it makes more mistakes on images containing the *sky* concept ( Fig. 6(a)(b) ). Based on such observation, the user hypothesizes that combining DenseNet and ResNet-101 may result in a stronger model. To verify such a hypothesis, we construct a simple ensemble model which takes the prediction (in the form of class probability) from both DenseNet and ResNet-101 and average the results to obtain the final class prediction. We further compare the ensemble model with DenseNet and ResNet-101 and observe that it indeed corrects the miss-classification of both models ( Fig. 6(c)(d) ). We further verify the results by comparing the overall accuracy on the ten classes and found that the ensemble model achieves 81.8% accuracy that outperforms both DenseNet (80.0%) and ResNet-101 (80.5%).

## 7.2 FCN and Semantic Segmentation

In this section, we will focus on presenting the insights discovered by ConceptExtract when analyzing an image semantic segmentation model for detecting unexpected objects on the road, usually lost cargos. The model is trained and tested on the public lost cargo dataset [33]. By utilizing our approach, we show that the model designers can obtain concepts that are both customized and human understandable. They can further utilize the insights generated from the concept to diagnose the model and improve model performance.

The lost cargo challenge addresses the problem of detecting unexpected small obstacles on the road often caused by lost cargo. To achieve this goal, a Fully Convolutional Network (FCN) [27] with a DenseNet Encoder [22] and Pyramid Scene Parsing [55] ( Fig. 3(B) ) is trained. We denote the model as DenseNet-FCN in our study.

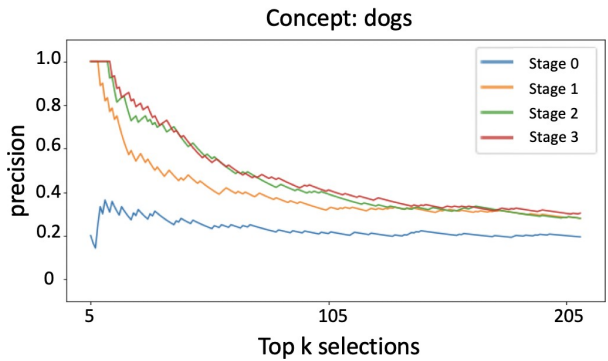

Fig. 7. The precision curves of the "dogs" concept extractor network in the lost cargo challenge. The fine annotations available in the dataset are used as the groundtruth. The top $k$ selections are made based on the concept confidence score. From the initial stage 0 to the final stage 3, we observe a significant improvement in the precision value especially for the top selections, validating the effectiveness of the active learning process and the usability of the concept extractor network.

edge cases (like object under shadows) where the model frequently makes mistakes. The visual concepts collected provide a good way to cluster the edge cases, reason about them, and develop corresponding mitigation strategies (such as adding artificial shadow augmentations).

## 8 EXPERIMENTAL VALIDATION

In an active learning process, the performance of the model varies with the training strategies. For example, even if two models are identical in structure, different methods for selecting the labeling candidate can generate significantly different model states. Another similar option is how much data should be labeled in each stage of the active learning process. In order to understand the impact of these variations, we carried out a series of experiments to evaluate the effectiveness of our active learning model in identifying these concepts under different settings. In the experiments, we mainly consider two main factors: *sampling strategy* and *latent representation*. The first one mainly refers to how we select the samples for the user to label at each stage. We compared concept extractors trained using different sampling methods and explained what sampling strategy we choose for ConceptExtract. For the second one, to investigate the influence of different latent presentations, besides the current DenseNet-FCN model, we used another well-trained model—VGG-16 to extract the latent representations of the same data and trained the active learning models for the same concepts. We compared the accuracy of these concept extractors and demonstrated the concepts these extractors generate. Finally, we also make justifications about other choices in ConceptExtract including model architecture, cutout methods, and prioritizing methods.

### 8.1 Concept Quality vs. Sampling Strategy

One of the main features of ConceptExtract is that it can include human knowledge in the training process of the active learning model. The users are able to choose which image patches to add to the training set for each stage. Another feature to involve human knowledge is that the user can brush on the image patches to mark the pixels containing the concept of their interest. To study if these two kinds of human knowledge can actually lead to a better concept extractor, we set up a *baseline model*. The baseline model shares the same network architecture as the active learning model and it is also trained in 4 stages. For each stage, 10,4,4,4 positive and negative images are needed respectively (For convenience, we use "x-x-x-x" to refer to the number of samples added to each training stage). All the processes like the data augmentation and the training epochs are the same except that instead of using the images chosen and brushed by users, the baseline model mimics a standard active learning approach where the most confused images will be labeled and added to the next training stage. To model a fully-labeled setting, we also create an *upper bound model*, using all image patches as training data and training the model in a single stage.

| Model | road(%) | lost cargo(%) | others(%) |
|---|---|---|---|
| **DenseNet-FCN-B** | 75.8 | 50.6 | 95.4 |
| **DenseNet-FCN-SA** | 83.6 (0.1) | 53.1 (0.5) | 96.8 (0.0) |
| **DeepLabV3+-B** | 82.8 | 57.4 | 96.8 |
| **DeepLabV3+-SA** | 82.8 (0.5) | 58.6 (0.3) | 96.8 (0.1) |

Table 1. Lost cargo segmentation performance after training on data with shadow augmentation. *-B are baseline models trained without shadow augmentation and *-SA are models trained with shadow augmentations. Numbers in parentheses are standard deviations. The table shows IoU accuracy for each semantic segmentation class.

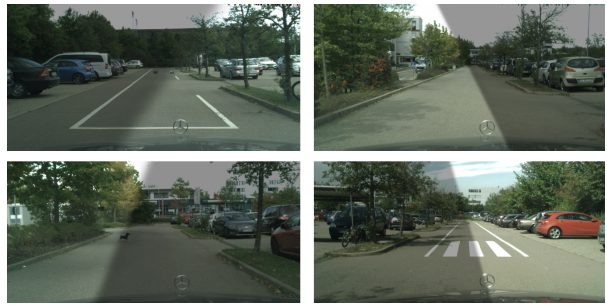

Fig. 8. This figure shows sample images from augmenting the training set with artificially-generated shadows (Section 7.2). The correlation of the shadow concept with misclassifications of the FCN model led to us augmenting the training distribution. The fine-tuned FCN model after augmentation is more accurate, validating the importance of shadow. Even though the shadow we generated is not realistic, fine-tuning still provides a substantial performance improvement.

An important choice to have to make during the active learning process is how many positive and negative samples should be labeled by the user at each stage. We seek a sampling strategy that makes the concept extraction network as accurate as possible while not requiring too many manually labeled samples. We tried a number of combinations in labeled images from each stage.

In Fig. 9, we demonstrate the precision curves of the concept extractors for "standard objects" and "bobby cars" under different sampling methods. The five models are: (i) baseline model, the model trained without human knowledge; (ii) 2-2-2-2, for each stage, two positive and negative images are added to the training set; (iii) 6-4-4-4, for the initialization, six positive and negative images are added; for the other three stages, four positive and four negative images are added each time; (iv) 10-4-4-4, which is also the current model used in ConceptExtract; (v) upper bound model, trained using all the data in one stage. As shown in the figure, all active learning models perform better than the baseline model.

This result illustrates that including human knowledge can improve the quality of the concept extractor. As we increase the number of training points for each stage, we see obvious precision improvements for the concept extractors as well. This suggests that we should include more image patches in the training process, especially for the first stage. Comparing these two concepts, "standard objects" are relatively easier to identify; using only about 3% of the data, we obtain a good concept extractor (at 80% accuracy for the top 100 selections). Although performance is not as good as the upper bound model as we expand the selections, the concept extractor still has a stable precision value. On the other hand, "bobby cars" is a more difficult concept to extract. Even using all the data, the top selection after 50 is lower than 0.5. Nevertheless, using only 7% of the positive data, we have already obtained a concept extractor approaching near upper bound performance, especially within the most confident predictions.

### 8.2 Concept Quality vs. Latent Representation

In Section 8.1, we have seen different degrees of difficulty for training different concept extractors. We also find that the overall performance for the concept extractor network of "standard objects" is better than the one of "bobby cars." We mainly have two possible hypotheses. One assumption is about the model architecture. Since we are using a shallow network for short training time, *is the active learning model complex enough to identify different concepts?* Another assumption is about the latent representation. Since the extractors are using the deep embeddings extracted from the task model as input instead of the image's pixels, *will the deep embeddings be capable of capturing these concept features in the latent space?* To investigate how different deep embeddings will affect the quality of the extractors, we carried out an experiment, where we obtained a pre-trained VGG-16 model from Pytorch, sent the image patches through this model and extracted a new set of deep embeddings. After that, we trained the extractors for the same concepts using these new deep embeddings. Comparing to the image segmentation model (FCN) we originally used, this VGG-16 model is well-trained and the performance for the classification task is fairly good.

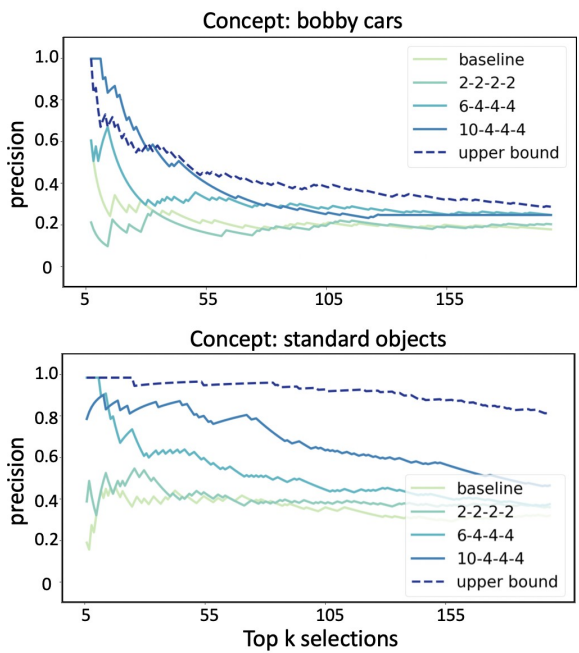

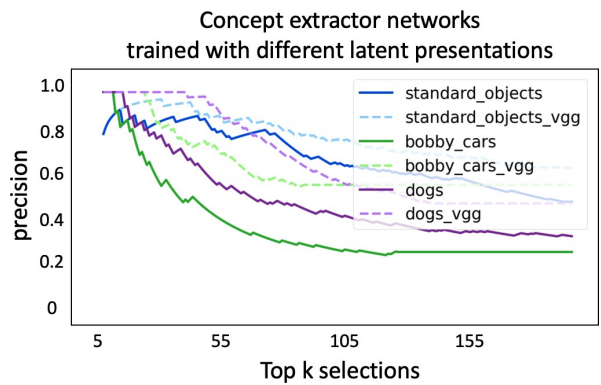

Fig. 9. The precision curves (plotted based on top-k selections based on the concept confidence score) for two concept extractors: bobby cars and standard objects under five sampling strategies. "baseline" is the model trained without human knowledge; "2-2-2-2" means for each stage, two positive images, and two negative images are added to training set; similar to "6-4-4-4"; the current model is the one used in the system, using a 10-4-4-4 sampling strategy; the upper bound model trains all the data in one stage. The curves show that the quality of the extractors for different concepts can vary significantly. This is mainly caused by the latent representations we use, which we discuss in Section 8.2.

Fig. 10. The precision curves for three concept extractors (standard objects, bobby cars, dogs) trained with two different deep embeddings: VGG-16 and FCN. The VGG-16 model is better trained than the FCN model and the concept extractors using deep embeddings from VGG-16 (dotted lines) outperform the ones from the FCN model (solid lines).

In Fig. 10, we show results for "standard objects", "bobby cars", and "dogs", concepts trained with two different deep embeddings. For all concepts, extractors using embeddings of VGG-16 perform better than those of FCN. For "bobby cars" and "dogs", the difference is substantial: precision values are roughly 0.7 and 0.9 for the top 50 selections, compared to those using the FCN model, which is about 0.4 and 0.5.

This experiment also verifies that even though the architecture of our concept extraction network is very simple, it can still extract concepts from the image patches. It also indicates that we can relate the extraction network's inability to extract a concept to the quality of the task model representations in the latent space. For an image segmentation model like FCN, since the model is trained on the specific task to differentiate lost cargo, road and backgrounds, it is not able to distinguish different type of cargoes like bobby cars and dogs, or telling humans and trees from the background very well. In contrast, a high-quality large image classifier like VGG-16 readily separates these features, providing evidence that such model understands the concepts being extracted.

### 8.3 Other Considerations

**How complex must the concept learning model be?** We ran an ablation study where we increased and decreased model complications a little by adding or removing hidden layers. Additional layers generally make convergence harder, and removing layers lowers the model accuracy significantly (see supplemental materials).

**How should we prioritize the display of active learning candidate images?** Instead of the common practice of letting users focus on images where the active learning model is least confident, we instead focus on identifying confident but wrong images from the perspective of the base model. This choice maximizes the performance of the concept model in its most confident predictions, which matches the goal of ConceptExtract better: extracting human understandable concepts via exemplar images.

**How should users localize class information in images?** There are two natural candidate UI techniques: rectangular brushing or lassoing [30]. We hypothesized that lassoing a detailed outline of the concept instead of a rectangle would increase the accuracy of the concept extractor. After running an experiment on FCN and VGG-16, we saw no significant difference in model performance for these two methods. Thus we choose the simpler UI: brushing.

## 9 LIMITATIONS, FUTURE WORK, AND CONCLUSION

Although we have shown that our prototype system can outperform standard active learning approaches (the baseline model), there are still a number of limitations to be addressed in the future. Most importantly, we would like to understand specifically what parts of the interface provide the most benefit for the human-in-the-loop approach? For example, how important is the exploratory clustering we provide during the active learning process? How do different modules affect the choice the user makes in selecting image patches? Would a different image patch layout help? These questions also will probe our understanding of the role of human expertise in the process.

Secondly, the current system supports a limited number of training data and image patches, and the user interface choices we made would not be effective under a larger number of classes and image samples. Although it is always possible to sample a small number of images from the training set to present to a user, a full study of the impact of choosing to show more (or fewer) images remains necessary.

Finally, a full evaluation in real-world application scenarios remains needed to complement our preliminary expert evaluation. For future machine learning interpretability practice, we would like to study more examples about how machine learning experts use these concept-based explanations to improve and understand their models.

In conclusion, we presented a novel approach for extracting concepts to help interpret neural networks, and contributed the use of a visualization-assisted active learning loop to extract interpretable concepts. We integrate the full pipeline into an interactive visualization system, ConceptExtract. Through case studies and experimental validations, we show our approach can extract concepts that are both human understandable and customizable based on the user's interest.